\documentclass[a4paper,ngerman,twoside,notitlepage]{article}
\usepackage{float}
\usepackage{makeidx}
\usepackage{amsmath}

\input{tcilatex}
\newcommand{\vekt}[1]{\mbox{\boldmath $#1$\unboldmath}}

\begin{document}

\title{Problem with the Derivation \\
of the Navier-Stokes Equation\\
by Means of Zwanzig-Mori Projection Technique \\
of Statistical Mechanics}
\author{J. Piest \\
Meisenweg 13\\
D-24147 Klausdorf, Germany\\
piestj@aol.com}
\maketitle

\begin{abstract}
The derivation of the Navier-Stokes equation starting from the Liouville
equation using projection techniques yields a friction term which is
nonlinear in the \ velocity. Using the \ results of multilinear
mode-coupling technique for calculating equlibrium correlation functions, \
it is shown that the second-order part of the term is non-zero, thus leading
to \ an incorrect formula for the equation.

\end{abstract}

\section{Introduction}

The derivation of hydrodynamic equations by Zwanzig-Mori projection
technique is a well-established method of statistical mechanics; see, e. g.,
the presentations in \cite{gra}, \cite{zu}. But analyzing the results
further leads to a problem which is connected to the fact that the
Navier-Stokes equation is of second order in the velocity. By derivation,
the friction term is nonlinear in the velocity; in order to keep the
equation correct it would be necessary for the second-order part of the term
to vanish. But, as is shown in this paper, the derivation furnishes an
expression which is generally non-zero. Since the Navier-Stokes equation is
obviously a correct formula, the question arises which features of the
derivation process provide these undesired details.

\section{Some basic formulas}

In this section some definitions and formulas are listed which are found in
the textbooks of statistical mechanics. We are working with a simple fluid,
i. e. we are looking at a system of $N$\ point particles with positions $%
\mathbf{y}_{i}$\ and velocities $\mathbf{v}_{i}$\ which collect to the phase
space variable $z$. Indices which number the particles are denoted by $j$, $%
k,\cdots $\ . The space densities of the conserved variables mass, energy
and momentum, called $n,\;e,\;\mathbf{p}$, are collected to a matrix $a$:

\begin{equation}
a_{\alpha }(\mathbf{x})=\sum_{j=1}^{N}\tilde{a}_{j\alpha }(z)\delta (\mathbf{%
x}-\mathbf{y}_{j})  \label{2.1}
\end{equation}

\begin{equation}
\tilde{a}_{j\alpha }=\left( m,\,\frac{m}{2}\mathbf{v}_{j}^{2}+\frac{1}{2}%
\sum_{l\neq j}\varphi (|\mathbf{y}_{j}-\mathbf{y}_{l}|),\,m\,v_{ja}\right)
\label{2.2}
\end{equation}

The greek index $\alpha $\ runs over $1,\,2,\,a$\ where $a$\ runs from 1 to
3 so that $\alpha $\ attains 5 scalar values in total. $m$\ is the particle
mass, $\varphi (|\mathbf{y}_{j}-\mathbf{y}_{l}|)$\ the intermolecular
potential which I write $\varphi _{jl}$\ for short. The motion of the
particle system is described by the variable $z$\ as a function of time $t$
. From this, all phase space functions, especially the variables $a$, are
functions of time described by the Liouville equation:

\begin{equation}
\frac{da}{dt}=\mathcal{L}a  \label{2.3}
\end{equation}%
\begin{equation}
\mathcal{L=}\frac{1}{m}\sum_{j=1}^{N}(\nabla _{v_{j}}H\cdot \nabla
_{y_{j}}-\nabla _{y_{j}}H\cdot \nabla _{v_{j}})  \label{2.4}
\end{equation}

The Liouville Operator $\mathcal{L}$\ is presented here as a real
differential operator. $H$\ is the Hamilton function:

\begin{equation}
H=\sum_{j=1}^{N}\left( \frac{1}{2}m\mathbf{v}_{j}^{2}+\frac{1}{2}\sum_{l\neq
j}\varphi _{jl}\right)  \label{2.5}
\end{equation}

The formal solution of the Liouville equation reads:

\begin{equation}
a(t)=\func{e}^{i\mathcal{L}t}a(0)  \label{2.6}
\end{equation}

The conservation property of the $a$\ is expressed by:

\begin{equation}
\frac{da_{\alpha }}{dt}=-\nabla \cdot s=-\nabla _{c}s_{\alpha c}  \label{2.7}
\end{equation}

$s$\ are the flux densities. Paired indices are summed over. The $s$\ show a
formal structure similar to the $a$; cf. (\ref{2.1}). The momentum flux
density reads:

\begin{equation}
s_{ab}(\mathbf{x})=\sum_{j}\left( mv_{ja}v_{jb}-\frac{1}{2}\sum_{l\neq j}%
\frac{d\varphi }{dr}\mid _{\left| r_{jl}\right| }\frac{r_{jla}r_{jlb}}{%
\left| r_{jl}\right| }\right) \delta (\mathbf{x}-\mathbf{y}_{j})  \label{2.8}
\end{equation}

Here $\mathbf{r}_{jl}=\mathbf{y}_{j}-\mathbf{y}_{l}$ . - The expectation of
a phase space function $A(z,N)$ with respect to the probability density $%
f(z,N)$\ is defined:

\begin{equation}
\left\langle A\right\rangle =tr\left\{ A(z,N)f(z,N)\right\}  \label{2.9}
\end{equation}

The definition of the Operation $tr\{\}$\ reads generally:

\begin{equation}
tr\left\{ A(z,N)\right\} =\sum_{N=1}^{\infty }\int dz\,A(z,N)  \label{2.10}
\end{equation}

\section{Hydrodynamic equations}

\setcounter{equation}{0}

For the projection calculations the formalism in \cite{gra}\ is adopted
where the time dependence of the dynamic variables is considered while the
expectiations are defined with respect to the probability density at time $%
t=0$ (Heisenberg picture). Moreover, in exponential operators where the
projector $\mathcal{P}$\ and the Operator $\mathcal{L}$\ appear together in
the exponent, $\mathcal{L}$\ appears 'before' (left of) $\mathcal{P}$. In
this paper, this is called the $\mathcal{L}$-$\mathcal{P}$ formulation which
stems from Zwanzig (see \cite{zw} where the Schr\"{o}dinger picture is used;
the succession of Operators reverses if one switches to the Heisenberg
picture) while Mori \cite{mo}\ seems to prefer the $\mathcal{P}$-$\mathcal{L}
$\ formulation.

The result of the derivation in \cite{gra}\ are the formulas (8.1.13)\
together with (8.1.12), (8.1.7)\ there; in the denotation here:

\begin{equation}
\frac{d\langle a_{\alpha }\rangle }{dt}=-\nabla _{c}\langle s_{\alpha
c}\rangle _{L,t}+D_{\alpha }(t)  \label{3.1}
\end{equation}

The expression for the dissipative force reads:

\begin{equation}
D_{\alpha }(\mathbf{x},t)=-\nabla _{c}\int_{0}^{t}dt^{\prime }\int d\mathbf{x%
}^{\prime }R_{\alpha \beta cd}(\mathbf{x},t,\mathbf{x}^{\prime },t^{\prime
})\nabla _{d}^{\prime }b_{\beta }(\mathbf{x}^{\prime },t^{\prime })
\label{3.2}
\end{equation}

with the kernel function:

\begin{equation}
R_{\alpha \beta cd}(\mathbf{x},t,\mathbf{x}^{\prime },t^{\prime })=\langle 
\mathcal{[G}(t^{\prime },t)\hat{s}_{\alpha c}(\mathbf{x},t)]\hat{s}_{\beta
d}(\mathbf{x}^{\prime },t^{\prime })\rangle _{L,t^{\prime }}  \label{3.3}
\end{equation}

$\langle \rangle _{L,t}$ denotes the expectation with respect to the
probability density of the local equilibrium $f_{L}(t)$:

\begin{equation}
f_{L}(t)=\psi (N)\exp (\Psi (t)-a(z)\ast b(t))  \label{3.4}
\end{equation}

where%
\begin{equation}
\psi (N)=\frac{1}{N!}(\frac{m}{h})^{3N}  \label{3.5}
\end{equation}

$h$ is Planck's constant. $\ast $\ denotes an operation which consists of a
product, a summation over a greek index and a space integration; the
respective variables and indices are not written down. The conjugated
parameters $b_{\beta }(\mathbf{x},t)$\ are:

\begin{equation}
b=\{\beta (\frac{m}{2}u^{2}-\mu ),\beta ,-\beta \mathbf{u\}}  \label{3.6}
\end{equation}

here $\beta =1/(k_{B}T)$; $k_{B}$\ is Boltzmann's constant and $T$\ the
temperature; these as well as the chemical potential $\mu $\ und the fluid
velocity $\mathbf{u}$\ are generally functions of $\mathbf{x}$\ and $t$. $%
\Psi (t)$\ is a normative quantity which ensures $\func{tr}\{f_{L}(t)\}=1$.
In (\ref{3.3}), $\hat{s}$\ denotes the reduced flux densities:

\begin{equation}
\hat{s}(\mathbf{x},t)=(1-\mathcal{P}(t))s(\mathbf{x})  \label{3.9}
\end{equation}

$\mathcal{P}$\ is the Zwanzig-Mori projection operator; for any phase space
funktion $g$, it is defined:

\begin{equation}
\mathcal{P}g=\langle g\rangle _{L}+\langle g\,\delta a\rangle _{L}\ast
\langle \delta a\,\delta a\rangle _{L}^{-1}\ast \delta a  \label{3.10}
\end{equation}

$\delta a=a-\langle a\rangle _{L}$. $\langle \rangle ^{-1}$ denotes the
inverse matrix. Generally \ $\langle \rangle _{L}$, and therefore $\mathcal{P%
}$\ , depend on time.\ We have $R=0$\ for $\alpha =1$\ or/and \ $\beta =1$,
since then the reduced fluxes are zero. $\mathcal{G}(t^{\prime },t)$\ is a
time-ordered exponential operator:

\begin{equation}
\mathcal{G}(t^{\prime },t)=\exp _{-}\{\int_{t^{\prime }}^{t}dt^{\prime
\prime }\,\mathcal{L}(1-\mathcal{P}(t^{\prime \prime }))\}  \label{3.11}
\end{equation}

We need formula (\ref{3.1}) for $\alpha =1,a$. One obtains (\cite{gra}\
(8.4.1), (8.3.12) und (8.4.12)):

\begin{equation}
\langle s_{1c}\rangle _{L,t}=\rho u_{c}  \label{3.7}
\end{equation}%
\begin{equation}
\langle s_{ac}\rangle _{L,t}=\rho u_{a}u_{c}+P\delta _{ac}  \label{3.8}
\end{equation}

$\rho $\ is the macroscopic mass density, and $P$\ the pressure. - \ From
here on the processes considered are restricted. The Problem mentioned in
the introduction appears even under this specialization, and the necessary
formulas are considerably simplified. Constant mass density/temperature
processes are considered; then, the chemical potential $\mu $\ ist constant
too. From (\ref{3.6}), the term $\beta =2$\ in (\ref{3.2})\ is zero; thus,
the sum runs over $\beta =b$\ only. Finally, I confine to stationary
currents, i. e. $\mathbf{u=}\func{const}(t)$. Then, expectations of single
phase space functions as well as the conjugated parameters and $\mathcal{P}$%
\ are constant in time; $\mathcal{G}$\ reduces to a non-ordered exponential
operator. Finally, the upper limit of the time integral in (\ref{3.2})\ may
be extended to infinity. Then it is reasonable to take the time integral
under the definition of the kernel function. For $\alpha =a$, one obtains,
instead of (\ref{3.2}), (\ref{3.3}):

\begin{equation}
D_{a}(\mathbf{x},t)=\nabla _{c}\int d^{\prime }\mathbf{x\,}R_{abcd}(\mathbf{x%
},\mathbf{x}^{\prime })\nabla _{d}^{\prime }u_{b}(\mathbf{x}^{\prime })
\label{3.12}
\end{equation}

\begin{equation}
R_{abcd}(\mathbf{x},\mathbf{x}^{\prime })=\beta \int_{0}^{\infty }dt\langle 
\func{e}^{\mathcal{L}(1-\mathcal{P)}t}\hat{s}_{ac}(\mathbf{x})]\hat{s}_{bd}(%
\mathbf{x}^{\prime })\rangle _{L}  \label{3.13}
\end{equation}

Finally, from (\ref{3.1}), \ together with (\ref{3.7}), (\ref{3.8}), one
obtains for stationary constant density/temperature currents

\begin{equation}
\nabla \cdot \mathbf{u}=0  \label{3.14}
\end{equation}%
\begin{equation}
\rho u_{c}\nabla _{c}u_{a}=-\nabla _{a}P+D_{a}  \label{3.15}
\end{equation}

with (\ref{3.12}), (\ref{3.13}) for $D$. These are the incompressibility
condition and the momentum equation for stationary constant density and
temperature current.

$R$ and $D$\ are nonlinear functionals of the velocity field. From (\ref%
{3.12})\ it is obvious that in lowest order $D$\ is linear in $\mathbf{u}$.
One obtains this order when one takes $R$\ at $\mathbf{u}=0$. $\mathbf{u}$\
enters $R$\ exclusively via the formula for the local equilibrium (\ref{3.4}%
). We have:%
\begin{equation}
f_{L}(z,N)|_{\mathbf{u}=0}=f_{0}(z,N)  \label{3.16}
\end{equation}

$f_{0}$\ is the probability density of the thermodynamic equilibrium for
prescribed mass density and temperature:%
\begin{equation}
f_{0}(z,N)=\psi (N)\exp (\Psi _{0}+\beta (\mu N-H(z)))  \label{3.17}
\end{equation}

$\Psi _{0}$\ is the norming quantity corresponding to $\Psi $. $R$\ at $%
\mathbf{u}=0$\ is called $R^{(0)}$:

\begin{equation}
R_{abcd}^{(0)}(\mathbf{x}-\mathbf{x}^{\prime })=\beta \int_{0}^{\infty
}dt\langle \func{e}^{\mathcal{L}(1-\mathcal{P}_{0}\mathcal{)}t}(\hat{s}%
_{0})_{\alpha c}(\mathbf{x})](\hat{s}_{0})_{bd}(\mathbf{x}^{\prime })\rangle
_{0}  \label{3.18}
\end{equation}%
\begin{equation}
(\hat{s}_{0})_{\alpha c}=(1-\mathcal{P}_{0})s_{\alpha c}  \label{3.18a}
\end{equation}

\begin{equation}
\mathcal{P}_{0}g=\langle g\rangle _{0}+\langle g\,\delta _{0}a\rangle
_{0}\ast \langle \delta _{0}a\,\delta _{0}a\rangle _{0}^{-1}\ast \delta _{0}a
\label{3.19}
\end{equation}

$\langle \rangle _{0}$\ is the expectation with respect to total
equilibrium, and we have $\delta _{0}a=a-\langle a\rangle _{0}$. Since
correlations in total equilibrium are transitive, $R^{(0)}$\ depends on the
single space variable $\mathbf{x}-\mathbf{x}^{\prime }$\ only. Provided the
space integral over $R^{(0)}$\ exists, formula (\ref{3.12}), in the linear
approximation, can be localized. The space integral resembles the quantity \
in \cite{gra}, Gl. (8.5.20); though the formula is obtained there in a
somewhat different way. The integral obeys several symmetries, which lead to
the expression \cite{gra}\ (8.5.21); in the denotation used here ($D^{(1)}$
is the linear part of $D$):%
\begin{equation}
D_{a}^{(1)}(\mathbf{x})=\tilde{\gamma}_{abcd}\nabla _{c}\nabla _{d}u_{b}(%
\mathbf{x})  \label{3.20}
\end{equation}

\begin{equation}
\tilde{\gamma}_{abcd}=\int d\mathbf{x\,}R_{abcd}^{(0)}(\mathbf{x})=(\delta
_{ab}\delta _{cd}+\delta _{ad}\delta _{bc})\eta +\delta _{ac}\delta
_{bd}(\varsigma -\frac{2}{3}\eta )  \label{3.21}
\end{equation}

$\eta ,$ $\varsigma $\ are the (dynamical) shear and bulk viscosity. If this
is introduced into (\ref{3.20}), and (\ref{3.14})\ is allowed for, $%
\varsigma $\ eliminates, and (\ref{3.15})\ changes to:%
\begin{equation}
\rho u_{c}\nabla _{c}u_{a}=-\nabla _{a}P+\eta \nabla ^{2}u_{a}  \label{3.22}
\end{equation}

This is the Navier-Stokes equation f\"{u}r stationary constant
density/temperature flow.

\section{2$^{nd}$ order term of the friction force}

\setcounter{equation}{0}

The Navier-Stokes equation (\ref{3.22})\ is of second order in the velocity.
In order that it results correctly from the momentum equation (\ref{3.15})
as an approximation for small Reynolds numbers, the 2$^{nd}$ order part $%
D^{(2)}$\ of the friction force\ must vanish. This part is built in (\ref%
{3.12})\ with the linear part $R^{(1)}$\ of the kernel function (\ref{3.13}%
). By Taylor's theorem, the latter reads:%
\begin{equation}
R_{abcd}^{(1)}(\mathbf{x},\mathbf{x}^{\prime })=\int d\mathbf{x}^{\prime
\prime }\frac{\delta R_{abcd}(\mathbf{x},\mathbf{x}^{\prime })}{\delta u_{e}(%
\mathbf{x}^{\prime \prime })}|_{\mathbf{u}=0}\,u_{e}(\mathbf{x}^{\prime
\prime })  \label{5.1}
\end{equation}

It is useful to express the functional derivative with respect to $\mathbf{u}
$\ by the derivatives with respect to the conjugated parameters $b_{\epsilon
}$\ (\ref{3.6}). These are parametrically related to ; since $b_{1}$\ is of 2%
$^{nd}$ order in $\mathbf{u}$\ and $b_{2}=\func{const}(\mathbf{u})$, for $%
\mathbf{u=0}$\ only the index value $\epsilon =e$\ remains, and we have:%
\begin{equation}
\frac{\delta R_{abcd}(\mathbf{x},\mathbf{x}^{\prime })}{\delta u_{e}(\mathbf{%
x}^{\prime \prime })}|_{\mathbf{u=0}}=-\beta \frac{\delta R_{abcd}(\mathbf{x}%
,\mathbf{x}^{\prime })}{\delta b_{e}(\mathbf{x}^{\prime \prime })}|_{\mathbf{%
u=0}}  \label{5.2}
\end{equation}

The calculation of \ the functional derivative is performed in the appendix.
The result is (\ref{5.17}):

\begin{equation}
\frac{\delta R_{abcd}(\mathbf{x},\mathbf{x}^{\prime })}{\delta b_{e}(\mathbf{%
x}^{\prime \prime })}|_{\mathbf{u}=0}=-\beta \int_{0}^{\infty }dt\langle
\lbrack \func{e}^{(1-\mathcal{P}_{0})\mathcal{L}t}(\hat{s}_{0})_{ac}(\mathbf{%
x})](\hat{s}_{0})_{bd}(\mathbf{x}^{\prime })p_{e}(\mathbf{x}^{\prime \prime
})\rangle _{0}  \label{5.17a}
\end{equation}

With \ (\ref{5.1}), (\ref{5.2}), this yields:

\begin{eqnarray}
R_{abcd}^{(1)}(\mathbf{x},\mathbf{x}^{\prime }) &=&-\beta \int d\mathbf{x}%
^{\prime \prime }\frac{\delta R_{abcd}(\mathbf{x},\mathbf{x}^{\prime })}{%
\delta b_{e}(\mathbf{x}^{\prime \prime })}|_{\mathbf{u}=0}\,u_{e}(\mathbf{x}%
^{\prime \prime })  \notag \\
&=&\beta ^{2}\int d\mathbf{x}^{\prime \prime }\left( \int_{0}^{\infty
}dt\langle \lbrack \func{e}^{(1-\mathcal{P}_{0})\mathcal{L}t}(\hat{s}%
_{0})_{ac}(\mathbf{x})](\hat{s}_{0})_{bd}(\mathbf{x}^{\prime })p_{e}(\mathbf{%
x}^{\prime \prime })\rangle _{0}\right) u_{e}(\mathbf{x}^{\prime \prime })
\label{5.18}
\end{eqnarray}

Therefore, $D^{(2)}$ reads:

\begin{align}
& D_{a}^{(2)}(\mathbf{x},t)=  \notag \\
& =\beta ^{2}\nabla _{c}\int d\mathbf{x}^{\prime }\int d\mathbf{x}^{\prime
\prime }\left( \int_{0}^{\infty }dt\langle \lbrack \func{e}^{(1-\mathcal{P}%
_{0})\mathcal{L}t}(\hat{s}_{0})_{ac}(\mathbf{x})](\hat{s}_{0})_{bd}(\mathbf{x%
}^{\prime })p_{e}(\mathbf{x}^{\prime \prime })\rangle _{0}\right) u_{e}(%
\mathbf{x}^{\prime \prime })\nabla _{d}^{\prime }u_{b}(\mathbf{x}^{\prime })
\label{5.19}
\end{align}

\subsection{Calculation of the kernel function}

In this subsection, the calculation of the kernel function in (\ref{5.19})
(the quantity in parentheses) is performed, using a result of multilinear
mode-coupling theory. Instead of (\ref{2.1}), orthonormal phase space
densities are used. By replacing the energy density $e$ by a suitable linear
combination of $e$ and $n$, the densities can be made orthogonal:%
\begin{equation}
\int d\mathbf{x}\langle \delta _{0}a_{\alpha }(\mathbf{x})\delta
_{0}a_{\beta }^{\ast }(0)\rangle _{0}=\chi _{\alpha }\delta _{\alpha \beta }
\label{5.20}
\end{equation}

The orthonormal variables $h_{\alpha }$\ read: 
\begin{equation}
h_{\alpha }=\frac{\delta _{0}a_{\alpha }}{\sqrt{\chi _{\alpha }}}
\label{5.21}
\end{equation}

The flux densities conjugated to the $h_{\alpha }$\ are called $r_{\alpha c}$%
, the corresponding reducted quantities $\hat{r}_{\alpha c}=(1-\mathcal{P}%
_{0})r_{\alpha c}$. By introducing the new definitions into (\ref{5.19}) one
obtains:

\begin{align}
D_{a}^{(2)}(\mathbf{x})& =  \notag \\
& =\beta ^{\frac{1}{2}}\rho ^{\frac{3}{2}}\nabla _{c}\int d\mathbf{x}%
^{\prime }\int d\mathbf{x}^{\prime \prime }\left( \int_{0}^{\infty
}dt\langle \lbrack \func{e}^{(1-\mathcal{P}_{0})\mathcal{L}t}\hat{r}_{ac}(%
\mathbf{x})]\hat{r}_{bd}(\mathbf{x}^{\prime })p_{e}(\mathbf{x}^{\prime
\prime })\rangle _{0}\right) u_{e}(\mathbf{x}^{\prime \prime })\nabla
_{d}^{\prime }u_{b}(\mathbf{x}^{\prime })  \label{5.22}
\end{align}

The calculation is done in Fourer space. We have:

\begin{equation}
D_{a}^{(2)}(\mathbf{k})=\beta ^{\frac{1}{2}}\rho ^{\frac{3}{2}}\frac{1}{%
(2\pi )^{6}}\int d\mathbf{q\,}N_{abc}(\mathbf{k},\mathbf{q},\mathbf{k-q}%
)u_{b}(\mathbf{q})u_{c}(\mathbf{k-q})  \label{5.23}
\end{equation}

\begin{equation}
N_{abc}(\mathbf{k},\mathbf{q},\mathbf{q}^{\prime
})=k_{d}\,q_{e}\int_{0}^{\infty }dt\langle \lbrack \func{e}^{(1-\mathcal{P}%
_{0})\mathcal{L}t}\hat{r}_{ad}(\mathbf{k})]\hat{r}_{be}^{\ast }(\mathbf{q}%
)h_{c}^{\ast }(\mathbf{q}^{\prime })\rangle _{0}  \label{5.24}
\end{equation}

The kernel function $N$\ is to be calculated. The multilinear mode-coupling
technique is for calculating correlation functions which are defined with
the original exponential operator $\exp \{Lt\}$, that is, without $1-%
\mathcal{P}_{0}$. In order to find a connection, one uses the operator
identity:

\begin{equation}
\func{e}^{(1-\mathcal{P}_{0})\mathcal{L}t}=\func{e}^{\mathcal{L}%
t}-\int_{0}^{t}dt^{\prime }\func{e}^{\mathcal{L}t^{\prime }}\mathcal{PL}%
\func{e}^{(1-\mathcal{P}_{0})\mathcal{L(}t-t^{\prime })}  \label{5.25}
\end{equation}

Application results in the formula:

\begin{equation}
N_{abc}(\mathbf{k},\mathbf{q},\mathbf{q}^{\prime })=iq_{e}\int_{0}^{\infty
}dt\left\{ \int_{0}^{t}dt^{\prime }\,\tilde{\Gamma}_{\alpha \epsilon }(%
\mathbf{k},t-t^{\prime })\langle \lbrack \func{e}^{\mathcal{L}t^{\prime
}}h_{\epsilon }(\mathbf{k})]\hat{r}_{be}^{\ast }(\mathbf{q})h_{c}^{\ast }(%
\mathbf{q}^{\prime })\rangle -ik_{d}\langle \lbrack \func{e}^{\mathcal{L}t}%
\hat{r}_{ad}(\mathbf{k})]\hat{r}_{be}^{\ast }(\mathbf{q})h_{c}^{\ast }(%
\mathbf{q}^{\prime })\rangle \right\} \text{ \ \ \ \ \ \ \ \ \ \ \ \ \ \ \ \
\ \ \ \ \ \ \ \ \ \ \ \ \ \ }  \label{5.26}
\end{equation}

\begin{equation}
\tilde{\Gamma}_{\alpha \beta }(\mathbf{k},t)=k_{c}k_{d}\langle \tilde{\Phi}%
_{\alpha c}(t)\tilde{\Phi}_{\beta d}^{\ast }\rangle (\mathbf{k})|_{\mathbf{k}%
=0}  \label{5.27}
\end{equation}%
\begin{equation}
\tilde{\Phi}_{\alpha c}(\mathbf{k},t)=\func{e}^{(1-\mathcal{P}_{0})\mathcal{L%
}t}\hat{r}_{\alpha c}(\mathbf{k})  \label{5.28}
\end{equation}

$\langle A\,B^{\ast }\rangle (\mathbf{k})$\ is defined to be the Fourier
transform of the correlation $\langle A(\mathbf{x})B(0)\rangle $. $\tilde{%
\Gamma}_{\alpha \beta }$\ is the localized memory function of the process.
The \ $\widetilde{}$\ \ points to the fact that the function $\tilde{\Phi}$\
is defined with the aid of the projection operator $\mathcal{P}_{0}$\ which
is linear in the microscopic densities $h_{\alpha }$\ (see below for the
discussion of the projection operator of the multilinear technique). It is
assumed that the time integral of the memory function exists:

\begin{equation}
\tilde{\gamma}_{\alpha \beta }(\mathbf{k})=\int_{0}^{\infty }dt\,\tilde{%
\Gamma}_{\alpha \beta }(\mathbf{k},t)  \label{5.29}
\end{equation}

Then, $\tilde{\Gamma}$ in (\ref{5.26})\ can be ''localized'' in time. $%
\tilde{\gamma}_{\alpha \beta }$\ is found to be diagonal; thus, if $\alpha
=a $, then $\beta =b$. The other correlations appearing in (\ref{5.26}) can
be related to the triple correlation $C_{3}$\ of the orthonormal densities:

\begin{equation}
(C_{3})_{\alpha \beta \gamma }(t)=\langle \lbrack \func{e}^{\mathcal{L}%
t}h_{\alpha }]h_{\beta }^{\ast }h_{\gamma }^{\ast }\rangle  \label{5.30}
\end{equation}

Finally, for the application in (\ref{5.23})\ it is possible to replace $%
N_{abc}$\ by the symmetrized form $\frac{1}{2}(N_{abc}+N_{acb})$\ which is
denoted by the same symbol. One obtains from (\ref{5.26}):

\begin{eqnarray}
2N_{abc} &=&-\tilde{\kappa}_{a\epsilon }\left[ (C_{3})_{\epsilon
bc}(0)-\int_{0}^{\infty }dt(i\,\omega _{b\tau }(C_{3})_{\varepsilon \tau
c}(t)+i\,\omega _{c\tau }(C_{3})_{\varepsilon b\tau }(t))\right]  \notag \\
&&-\frac{\partial (C_{3})_{abc}(t)}{\partial t}|_{t=0}-i\,\omega _{b\tau
}(C_{3})_{a\tau c}(0)-i\,\omega _{c\tau }(C_{3})_{ab\tau }(0)  \label{5.31}
\end{eqnarray}%
\begin{equation}
\tilde{\kappa}_{\alpha \beta }=\tilde{\gamma}_{\alpha \beta }+i\,\omega
_{\alpha \beta }  \label{5.31a}
\end{equation}

\begin{equation}
\omega _{\alpha \beta }(\mathbf{k})=k_{d}\langle r_{\alpha d}\,h_{\beta
}^{\ast }\rangle (\mathbf{k})|_{\mathbf{k}=0}  \label{5.32}
\end{equation}

$N$ \ and $C_{3}$\ depend on three wave numbers \ $%
\vekt{k}%
$, $%
\vekt{q}%
$, $%
\vekt{q}%
^{\prime }$. The wave numbers of the other quantities in (\ref{5.31})\ are \ 
$\tilde{\gamma}_{a\epsilon }(\mathbf{k})$, $\omega _{a\epsilon }(\mathbf{k})$%
, $\omega _{b\tau }(\mathbf{q})$, $\omega _{c\tau }(\mathbf{q}^{\prime })$.

By (\ref{5.31}), we connected the kernel function $N$\ to the triple
correlation $C_{3}$\ which is defined with the exponential Operator $\exp
\{Lt\}$. The formula for the latter we take from the multilinear mode
coupling theory for correlation functions \cite{vzs}. This theory is worked
out in Fourier space \textit{before} performing the thermodynamic limit.
Then, the correlations are proportional to the Volume $V$. In what follows
all correlations are divided by $V$; then, $V$\ does not appear any more in
the formulas. From the formulas (31), (30) there, one obtains, with the
denotation used here:%
\begin{equation}
(C_{3})_{\alpha \beta \gamma }(t)=(C_{2})_{\alpha \epsilon }(t)J_{\epsilon
\beta \gamma }-\int_{0}^{t}dt^{\prime }(C_{2})_{\alpha \epsilon
}(t-t^{\prime })ik_{d}S_{\epsilon \rho \tau d}(C_{2})_{\rho \beta
}(t^{\prime })(C_{2})_{\tau \gamma }(t^{\prime })  \label{5.33}
\end{equation}%
\begin{equation}
(C_{2})_{\alpha \beta }(t)=\langle \lbrack \func{e}^{\mathcal{L}t}h_{\alpha
}]h_{\beta }^{\ast }\rangle  \label{5.34}
\end{equation}%
\begin{equation}
S_{\alpha \beta \gamma d}=\langle r_{\alpha d}h_{\beta }^{\ast }h_{\gamma
}^{\ast }\rangle |_{\mathbf{q},\mathbf{k},\mathbf{k}^{\prime }=0}
\label{5.35}
\end{equation}%
\begin{equation}
J_{\alpha \beta \gamma }=(C_{3})_{\alpha \beta \gamma }(0)|_{\mathbf{q},%
\mathbf{k},\mathbf{k}^{\prime }=0}  \label{5.36}
\end{equation}

For $C_{2}$, one obtains, after ''localization'' in time corresponding to (%
\ref{5.29}):

\begin{equation}
(\dot{C}_{2})_{\alpha \beta }(t)=-\kappa _{\alpha \gamma }(C_{2})_{\gamma
\beta }(t)  \label{5.37}
\end{equation}

\begin{equation}
\kappa _{\alpha \beta }=\gamma _{\alpha \beta }+i\,\omega _{\alpha \beta }
\label{5.38}
\end{equation}

$\gamma $\ is the dissipation matrix defined corresponding to (\ref{5.29})\
but with the projection operator of the multilinear Theory which is
nonlinear in the $h_{\alpha }$. When one introduces (\ref{5.33})\ into(\ref%
{5.31}) , one obtains an expression for $N$ which contains the matrices $%
\omega $, $\gamma $, \ $\tilde{\gamma}$, $J$\ and $S$. For the present
intention to check whether the 2$^{nd}$ order part of the friction force is
different from zero, one can neglect the difference between $\gamma $\ and $%
\tilde{\gamma}$. Then one finds:

\begin{equation}
N_{abc}=\frac{1}{2}(\gamma _{b\rho }\delta _{c\sigma }+\delta _{b\rho
}\gamma _{c\sigma })(\kappa _{\rho \epsilon }\delta _{\sigma \tau }+\delta
_{\rho \epsilon }\kappa _{\sigma \tau })^{-1}ik_{d}S_{a\epsilon \tau d}
\label{5.39}
\end{equation}

The kernel function $N$\ will be used to calculate the friction force term $%
D_{2}$\ (\ref{5.23}), with the incompressibility condition (\ref{3.14}) $%
\mathbf{k}\cdot \mathbf{u}=0$\ to be incorporated. This being accounted for,
for $N$\ there remains a relevant part:%
\begin{equation}
N_{abc}=\frac{1}{2}ik_{d}S_{abcd}  \label{5.40}
\end{equation}

One finds:%
\begin{equation}
S_{abcd}=\frac{1}{\sqrt{\rho \beta }}(\delta _{ab}\delta _{cd}+\delta
_{ac}\delta _{bd}-\lambda \delta _{ad}\delta _{bc})  \label{5.41}
\end{equation}%
\begin{equation}
\lambda =\frac{\partial P}{\partial \beta }|\rho (\frac{\partial \epsilon }{%
\partial \beta }|\rho )^{-1}  \label{5.42}
\end{equation}

and by introduction into (\ref{5.23}):

\begin{equation}
D_{a}^{(2)}(\mathbf{k})=\frac{\rho }{(2\pi )^{6}}\int d\mathbf{q}\left(
ik_{d}\,u_{a}(\mathbf{q})u_{d}(\mathbf{k}-\mathbf{q})-\frac{\lambda }{2}%
ik_{a}\,u_{b}(\mathbf{q})u_{b}(\mathbf{k}-\mathbf{q})\right)  \label{5.43}
\end{equation}

It is seen that the 2$^{nd}$ order part of the friction force is calculated
to be not zero. The formula will be evaluated in the next section.

\section{Consequences}

\setcounter{equation}{0}

The consequences of the result (\ref{5.43})\ will show up clearly if one
uses the solenoidal form of the Navier-Stokes equation, that is, the
equation with the pressure term elimited. Instead of (\ref{3.22}), the
instationary form of the equation is chosen, and the 2$^{nd}$ order term of
the friction force $D^{(2)}{}_{a}=\nabla _{c}R_{ac}^{(2)}$\ ist added. Here $%
R_{ac}$\ with two indices is the stress tensor which should not be
confounded with the kernel function $R_{abcd}$. The calculation is performed
in Fourier space. The Navier-Stokes equation supplemented by $D^{(2)}$
reads: 
\begin{equation}
\rho (\frac{\partial u_{a}}{\partial t}+\nu
k^{2}u_{a})=-ik_{a}P-ik_{c}\left( \frac{\rho }{(2\pi )^{3}}\int d\mathbf{q\,}%
u_{a}(\mathbf{q})u_{c}(\mathbf{k}-\mathbf{q})-R_{ac}^{(2)}\right)
\label{6.1}
\end{equation}

After a short calculation (see, e. g., \cite{mcb}\ appendix D, (D34)) the
solenoidal form of the equation is obtained:%
\begin{equation}
\rho (\frac{\partial u_{a}}{\partial t}+\nu k^{2}u_{a})=-ik_{c}\varepsilon
_{ab}\left( \frac{\rho }{(2\pi )^{3}}\int d\mathbf{q\,}u_{b}(\mathbf{q}%
)u_{c}(\mathbf{k}-\mathbf{q})-R_{bc}^{(2)}\right)  \label{6.2}
\end{equation}%
\begin{equation}
\varepsilon _{ab}=\delta _{ab}-\hat{k}_{a}\hat{k}_{b}  \label{6.3}
\end{equation}

$\widehat{\mathbf{k}}$\ is the unit vector attached to $\mathbf{k}$. Now,
for $D_{2}$\ the result (\ref{5.43})\ is introduced. Since we have $%
k_{c}\varepsilon _{ac}=0$, the 2$^{nd}$ term does not contribute. One
obtains:%
\begin{equation}
ik_{c}\varepsilon _{ab}\left( \frac{\rho }{(2\pi )^{3}}\int du_{b}(\mathbf{q}%
)u_{c}(\mathbf{k}-\mathbf{q})-R_{bc}^{(2)}\right) =0  \label{6.4}
\end{equation}

The 2$^{nd}$ order part of the friction force cancels the convolution term
of the Navier-Stokes equation, so that the equation reduces to its linar
part. Since by phenomenological evidence the convolution term is an
essential part of the equation, this theoretical result cannot be correct.

\section{Summary}

\setcounter{equation}{0}

The derivation of hydrodynamic equations by Zwanzig-Mori projection
technique has been reviewed. For simplicity, incompressible constant
density/temperature fluids have been considered. The friction force term in
the momentum equation is a nonlinear functional of the fluid velocity; when
the \ linear approximation is taken, the Navier-Stokes equation is obtained.
In the present paper, the second-order term of the friction force has been
calculated. It contains a three-point time correlation function which has
been evaluated using a result of multilinear mode-coupling theory;\ the
final result is (\ref{5.43}). Since the Navier-Stokes equation is second
order in the velocity, in order to obtain it properly as an approximation of
the momentum equation for small Reynolds numbers, one would expect the
second-order part of the friction force to vanish. Thus, (\ref{5.43}) is
clearly an undesired result; it would be very valuable to know which detail
of the derivation may be responsible for it.

\appendix

\section{Appendix: Calculation of the functional derivative}

\setcounter{equation}{0}

In order to calculate the derivative in (\ref{5.2}), the formula for $R$\ (%
\ref{3.13})\ is written in detail:%
\begin{equation}
R_{abcd}(\mathbf{x},\mathbf{x}^{\prime })=\beta \int_{0}^{\infty }dt\,\func{%
tr}\{f_{L}[\func{e}^{\mathcal{L}(1-\mathcal{P})t}(1-\mathcal{P})s_{ac}(%
\mathbf{x})](1-\mathcal{P})s_{bd}(\mathbf{x}^{\prime })\}  \label{5.3}
\end{equation}

The expression depends on $b_{e}$\ 4-fold, namely, in the formula for the
local equilibrium $f_{L}$, and in $\mathcal{P}$\ appearing 3-fold. For
abbreviation, the formula is written:%
\begin{equation}
\frac{\delta R_{abcd}(\mathbf{x},\mathbf{x}^{\prime })}{\delta b_{e}(\mathbf{%
x}^{\prime \prime })}=\sum_{i=1}^{4}\left[ \frac{\delta R}{\delta b}\right]
^{(i)}  \label{5.4}
\end{equation}

In the consecutive formulas, in the first row the definition of the several
parts is expressed. For the calculation some auxiliary theorems for
projection operators are used, which can be found in the text books. For the
1$^{st}$ partial term:%
\begin{eqnarray}
\left[ \frac{\delta R_{abcd}}{\delta b_{e}(\mathbf{x}^{\prime \prime })}%
\right] ^{(1)} &=&\beta \int_{0}^{\infty }dt\,\func{tr}\{\frac{\delta f_{L}}{%
\delta b_{e}(\mathbf{x}^{\prime \prime })}[\func{e}^{\mathcal{L}(1-\mathcal{P%
})t}\hat{s}_{ac}(\mathbf{x})]\hat{s}_{bd}(\mathbf{x}^{\prime })\}  \notag \\
&=&-\beta \int_{0}^{\infty }dt\langle \lbrack \func{e}^{\mathcal{L}(1-%
\mathcal{P})t}\hat{s}_{ac}(\mathbf{x})]\hat{s}_{bd}(\mathbf{x}^{\prime
})\delta p_{e}(\mathbf{x}^{\prime \prime })\rangle _{L}  \label{5.5}
\end{eqnarray}

The 2$^{nd}$ term:

\begin{eqnarray}
\left[ \frac{\delta R_{abcd}}{\delta b_{e}(\mathbf{x}^{\prime \prime })}%
\right] ^{(2)} &=&-\beta \int_{0}^{\infty }dt\langle \lbrack \func{e}^{%
\mathcal{L}(1-\mathcal{P})t}\hat{s}_{ac}(\mathbf{x})]\frac{\delta \mathcal{P}%
}{\delta b_{e}(\mathbf{x}^{\prime \prime })}s_{bd}(\mathbf{x}^{\prime
})\rangle _{L}  \notag \\
&=&\beta \int_{0}^{\infty }dt\langle \lbrack \func{e}^{\mathcal{L}(1-%
\mathcal{P})t}\hat{s}_{ac}(\mathbf{x})]\mathcal{P}\hat{s}_{bd}(\mathbf{x}%
^{\prime })\delta p_{e}(\mathbf{x}^{\prime \prime })\rangle _{L}  \label{5.6}
\end{eqnarray}

The Operator $\mathcal{P}$\ acts on everything on the right (provided the
action is not limited by parentheses). The 3$^{rd}$ term:%
\begin{eqnarray}
\left[ \frac{\delta R_{abcd}}{\delta b_{e}(\mathbf{x}^{\prime \prime })}%
\right] ^{(3)} &=&-\beta \int_{0}^{\infty }dt\langle \lbrack \func{e}^{%
\mathcal{L}(1-\mathcal{P})t}\frac{\delta \mathcal{P}}{\delta b_{e}(\mathbf{x}%
^{\prime \prime })}s_{ac}(\mathbf{x})]\hat{s}_{bd}(\mathbf{x}^{\prime
})\rangle _{L}  \notag \\
&=&\beta \int_{0}^{\infty }dt\langle \lbrack \func{e}^{\mathcal{L}(1-%
\mathcal{P})t}\mathcal{P}\hat{s}_{ac}(\mathbf{x})\delta p_{e}(\mathbf{x}%
^{\prime \prime })]\hat{s}_{bd}(\mathbf{x}^{\prime })\rangle _{L}  \notag \\
&=&\beta \int_{0}^{\infty }dt\langle \lbrack \func{e}^{(1-\mathcal{P})%
\mathcal{L}t}(1-\mathcal{P})\mathcal{P}\hat{s}_{ac}(\mathbf{x})\delta p_{e}(%
\mathbf{x}^{\prime \prime })]\hat{s}_{bd}(\mathbf{x}^{\prime })\rangle _{L}=0
\label{5.7}
\end{eqnarray}

The 4$^{th}$ term is defined:%
\begin{equation}
\left[ \frac{\delta R_{abcd}}{\delta b_{e}(\mathbf{x}^{\prime \prime })}%
\right] ^{(4)}=\beta \int_{0}^{\infty }dt\langle \lbrack \frac{\delta \func{e%
}^{\mathcal{L}(1-\mathcal{P})t}}{\delta b_{e}(\mathbf{x}^{\prime \prime })}%
\hat{s}_{ac}(\mathbf{x})]\hat{s}_{bd}(\mathbf{x}^{\prime })\rangle _{L}
\label{5.8}
\end{equation}

The formula for the derivative of the exponential operator reads:%
\begin{equation}
\frac{\delta \func{e}^{\mathcal{L}(1-\mathcal{P})t}}{\delta b_{e}(\mathbf{x}%
^{\prime \prime })}=\int_{0}^{t}dt^{\prime }\func{e}^{\mathcal{L}(1-\mathcal{%
P})t^{\prime }}\mathcal{LP}\delta p_{e}(\mathbf{x}^{\prime \prime })(1-%
\mathcal{P})\func{e}^{\mathcal{L}(1-\mathcal{P})(t-t^{\prime })}  \label{5.9}
\end{equation}

We use the identity:%
\begin{equation}
\int_{0}^{\infty }dt\int_{0}^{t}dt^{\prime }A(t^{\prime },t-t^{\prime
})=\int_{0}^{\infty }dt\int_{0}^{\infty }dt^{\prime }A(t^{\prime },t)
\label{5.10}
\end{equation}

One obtains:%
\begin{eqnarray}
\left[ \frac{\delta R_{abcd}}{\delta b_{e}(\mathbf{x}^{\prime \prime })}%
\right] ^{(4)} &=&\beta \int_{0}^{\infty }dt\int_{0}^{\infty }dt^{\prime
}\langle \lbrack \func{e}^{\mathcal{L}(1-\mathcal{P})t^{\prime }}\mathcal{LP}%
\delta p_{e}(\mathbf{x}^{\prime \prime })(1-\mathcal{P})\func{e}^{\mathcal{L}%
(1-\mathcal{P})t}\hat{s}_{ac}(\mathbf{x})]\hat{s}_{bd}(\mathbf{x}^{\prime
})\rangle _{L}  \notag \\
&=&\beta \int_{0}^{\infty }dt\int_{0}^{\infty }dt^{\prime }\langle \lbrack 
\func{e}^{(1-\mathcal{P})\mathcal{L}t^{\prime }}(1-\mathcal{P})\mathcal{LP}%
\delta p_{e}(\mathbf{x}^{\prime \prime })\func{e}^{(1-\mathcal{P})\mathcal{L}%
t}\hat{s}_{ac}(\mathbf{x})]\hat{s}_{bd}(\mathbf{x}^{\prime })\rangle _{L} 
\notag \\
&=&\beta \int_{0}^{\infty }dt\int_{0}^{\infty }dt^{\prime }\langle \lbrack 
\frac{d}{dt^{\prime }}\func{e}^{(1-\mathcal{P})\mathcal{L}t^{\prime }}%
\mathcal{P}\delta p_{e}(\mathbf{x}^{\prime \prime })\func{e}^{(1-\mathcal{P})%
\mathcal{L}t}\hat{s}_{ac}(\mathbf{x})]\hat{s}_{bd}(\mathbf{x}^{\prime
})\rangle _{L}  \label{5.11}
\end{eqnarray}

The integration over $t^{\prime }$\ is performed:%
\begin{eqnarray}
\left[ \frac{\delta R_{abcd}}{\delta b_{e}(\mathbf{x}^{\prime \prime })}%
\right] ^{(4)} &=&\beta \int_{0}^{\infty }dt\left\{ \lim_{t^{\prime
}\rightarrow \infty }\langle \lbrack \func{e}^{(1-\mathcal{P})\mathcal{L}%
t^{\prime }}\mathcal{P}\delta p_{e}(\mathbf{x}^{\prime \prime })\func{e}^{(1-%
\mathcal{P})\mathcal{L}t}\hat{s}_{ac}(\mathbf{x})]\hat{s}_{bd}(\mathbf{x}%
^{\prime })\rangle _{L}\right.  \notag \\
&&-\left. \langle \lbrack \mathcal{P}\delta p_{e}(\mathbf{x}^{\prime \prime
})\func{e}^{(1-\mathcal{P})\mathcal{L}t}\hat{s}_{ac}(\mathbf{x})]\hat{s}%
_{bd}(\mathbf{x}^{\prime })\rangle _{L}\right\}  \label{5.12}
\end{eqnarray}

In the 1$^{st}$ term, the operation $\mathcal{P}$\ is performed by (\ref%
{3.10}):%
\begin{eqnarray}
&&\beta \int_{0}^{\infty }dt\lim_{t^{\prime }\rightarrow \infty }\langle
\lbrack \func{e}^{(1-\mathcal{P})\mathcal{L}t^{\prime }}\mathcal{P}\delta
p_{e}(\mathbf{x}^{\prime \prime })\func{e}^{(1-\mathcal{P})\mathcal{L}t}%
\hat{s}_{ac}(\mathbf{x})]\hat{s}_{bd}(\mathbf{x}^{\prime })\rangle _{L} 
\notag \\
&=&\beta \int_{0}^{\infty }dt\lim_{t^{\prime }\rightarrow \infty }\langle 
\left[ \func{e}^{(1-\mathcal{P})\mathcal{L}t^{\prime }}\left\{ \langle
\delta p_{e}(\mathbf{x}^{\prime \prime })\func{e}^{(1-\mathcal{P})\mathcal{L}%
t}\hat{s}_{ac}(\mathbf{x})\rangle _{L}\right. \right.  \notag \\
&&\text{ \ \ \ \ \ \ \ \ \ \ \ \ \ \ \ \ \ \ \ \ \ \ \ \ \ \ \ \ \ \ }%
+\left. \left. \langle \delta p_{e}(\mathbf{x}^{\prime \prime })[\func{e}%
^{(1-\mathcal{P})\mathcal{L}t}\hat{s}_{ac}(\mathbf{x})]\delta a\rangle
_{L}\langle \delta a\,\delta a\rangle _{L}^{-1}\delta a\right\} \right] 
\hat{s}_{bd}(\mathbf{x}^{\prime })\rangle _{L}  \notag \\
&=&\beta \int_{0}^{\infty }dt\left\{ \langle \delta p_{e}(\mathbf{x}^{\prime
\prime })\func{e}^{(1-\mathcal{P})\mathcal{L}t}\hat{s}_{ac}(\mathbf{x}%
)\rangle _{L}\langle \hat{s}_{bd}(\mathbf{x}^{\prime })\rangle _{L}\right. 
\notag \\
&&\text{ \ \ \ \ \ \ \ \ \ }+\left. \langle \delta p_{e}(\mathbf{x}^{\prime
\prime })[\func{e}^{(1-\mathcal{P})\mathcal{L}t}\hat{s}_{ac}(\mathbf{x}%
)]\delta a\rangle _{L}\langle \delta a\,\delta a\rangle
_{L}^{-1}\lim_{t^{\prime }\rightarrow \infty }\langle \lbrack \func{e}^{(1-%
\mathcal{P})\mathcal{L}t^{\prime }}\delta a]\hat{s}_{bd}(\mathbf{x}^{\prime
})\rangle _{L}\right\}  \label{5.13}
\end{eqnarray}

Here, the 2$^{nd}$ factor of the 1$^{st}$ term vanishes, therefore the 1$%
^{st}$ term vanishes in total. All expressions are finely to be taken at $%
b=b_{0}$. Therefore, the limit $\lim_{t^{\prime }\rightarrow \infty }$\ of a
correlation function in thermodynamic equilibrium must be performed. One
assumes that this equates to the product of the expectations of the factor
functions. This assumption has been checked with the aid of the multilinear
mode coupling theory, and verified for the special cases appearing here:

\begin{equation}
\lim_{t^{\prime }\rightarrow \infty }\langle \lbrack \func{e}^{(1-\mathcal{P}%
_{0})\mathcal{L}t^{\prime }}\delta _{0}a](1-\mathcal{P}_{0})s_{bd}(\mathbf{x}%
^{\prime })\rangle _{0}=\langle \delta _{0}a\rangle _{0}\langle (1-\mathcal{P%
}_{0})s_{bd}(\mathbf{x}^{\prime })\rangle _{0}=0  \label{5.14}
\end{equation}

with $\mathcal{P}_{0}$\ by (\ref{3.19}). Therefore, the 2$^{nd}$ term in (%
\ref{5.13})\ is zero also, so this is true for the 1$^{st}$ term in (\ref%
{5.12})\ in total. Finally, the 2$^{nd}$ term in (\ref{5.12})\ vanishes,
since one factor starts with $\mathcal{P}$, the other with $1-\mathcal{P}$.
One obtains:%
\begin{equation}
\left[ \frac{\delta R_{abcd}}{\delta b_{e}(\mathbf{x}^{\prime \prime })}%
\right] ^{(4)}|_{\mathbf{u}=0}=0  \label{5.15}
\end{equation}

Thus, we finally find that the derivative (\ref{5.4}), after letting $%
\mathbf{u}=0$, will be equal to the sum of the first two terms of the r. h.
s.; with (\ref{5.5}), (\ref{5.6}):%
\begin{eqnarray}
&&\left[ \frac{\delta R_{abcd}}{\delta b_{e}(\mathbf{x}^{\prime \prime })}%
\right] ^{(1)}+\left[ \frac{\delta R_{abcd}}{\delta b_{e}(\mathbf{x}^{\prime
\prime })}\right] ^{(2)}  \notag \\
&=&-\beta \int_{0}^{\infty }dt\langle \lbrack \func{e}^{\mathcal{L}(1-%
\mathcal{P})t}\hat{s}_{ac}(\mathbf{x})](1-\mathcal{P})\hat{s}_{bd}(\mathbf{x}%
^{\prime })\delta p_{e}(\mathbf{x}^{\prime \prime })\rangle _{L}  \notag \\
&=&-\beta \int_{0}^{\infty }dt\langle \lbrack \func{e}^{(1-\mathcal{P})%
\mathcal{L}t}\hat{s}_{ac}(\mathbf{x})]\hat{s}_{bd}(\mathbf{x}^{\prime
})\delta p_{e}(\mathbf{x}^{\prime \prime })\rangle _{L}  \label{5.16}
\end{eqnarray}

One sees that the shortest expression results with the $\mathcal{P}$-before-$%
\mathcal{L}$ formulation. Finally, by taking $\mathbf{u}=0$:%
\begin{equation}
\frac{\delta R_{abcd}(\mathbf{x},\mathbf{x}^{\prime })}{\delta b_{e}(\mathbf{%
x}^{\prime \prime })}|_{\mathbf{u}=0}=-\beta \int_{0}^{\infty }dt\langle
\lbrack \func{e}^{(1-\mathcal{P}_{0})\mathcal{L}t}(\hat{s}_{0})_{ac}(\mathbf{%
x})](\hat{s}_{0})_{bd}(\mathbf{x}^{\prime })p_{e}(\mathbf{x}^{\prime \prime
})\rangle _{0}  \label{5.17}
\end{equation}

The last factor $p_{e}$\ does not show a $\delta $\ any more since $\langle
p_{e}\rangle _{0}=0$. - The calculation of the derivative has been performed
in $\mathcal{P}$-before-$\mathcal{L}$ formulation also. The in-between steps
look somewhat different; for instance, we have $[]^{(4)}\neq 0$. The final
result coincedes with (\ref{5.16}).


\begin{thebibliography}{9}
\bibitem{gra} H. Grabert: Projection operator techniques in nonequilibrium
statistical mechanics. Springer, Berlin, Heidelberg, New York (1982)

\bibitem{zu} D. Zubarev, V. Morozov, G. R\"{o}pke: Statistical mechanics of
nonequilibrium processes. Akademie Verlag, Berlin (1996)

\bibitem{zw} R. Zwanzig, J Chem. Phys. 33: 1338 (1960)

\bibitem{mo} H. Mori, Progr. Theor. Phys. 33: 423 (1965)

\bibitem{mcb} W. D. McComb, The Physics of Fluid Turbulence, Clarendon
Press, Oxford, Reprint 1994

\bibitem{vzs} R. van Zon, J. Schofield, Phys. Rev. E 65 (2001), 011106
\end{thebibliography}
\end{document}